\documentclass[aps,prc,twocolumn,showpacs,showkeys,nofootinbib,superscriptaddress]{revtex4-1}

\usepackage{bm}
\usepackage{longtable}
\usepackage{graphicx}
\usepackage{amsmath}
\usepackage{amsfonts}
\usepackage{amssymb}
\usepackage{natbib} 
\usepackage{setspace}
\usepackage{xspace}
\usepackage{cancel}
\usepackage{physics}
\usepackage{tensor,braket,color,bbm,slashed}

\usepackage{newtxtext,newtxmath}
\usepackage{mathrsfs}

\begin{document}
\title{
  \begin{flushright}
    \rightline{APCTP Pre2019-013, LFTC-19-8/46} 
  \end{flushright}
  Valence-quark distributions of pions and kaons in a nuclear medium}

\author{Parada~T.~P.~Hutauruk}
\email{parada.hutauruk@apctp.org}
\affiliation{Asia Pacific Center for Theoretical Physics, Pohang, Gyeongbuk 37673, Korea}

\author{J.~J.~Cobos-Mart\'{\i}nez}
\email{jcobos@fis.cinvestav.mx}
\affiliation{C\'atedra CONACyT, Departamento de F\'{\i}sica, Centro de Investigaci\'on
  y de Estudios Avanzados del Instituto Polit\'ecnico Nacional, Apartado Postal 14-740,
  07000, Ciudad de M\'exico, M\'exico}
\affiliation{Asia Pacific Center for Theoretical Physics, Pohang, Gyeongbuk 37673, Korea}

\author{Yongseok Oh}
\email{yohphy@knu.ac.kr}
\affiliation{Department of Physics, Kyungpook National University, Daegu 41566, Korea}
\affiliation{Asia Pacific Center for Theoretical Physics, Pohang, Gyeongbuk 37673, Korea}

\author{K.~Tsushima}
\email{kazuo.tsushima@gmail.com}
\affiliation{Laborat\'orio de F\'isica Te\'orica e Computacional, 
Universidade Cruzeiro do Sul  / Universidade Cidade de Sao Paulo,
01506-000, S\~ao Paolo, SP, Brazil}
\affiliation{Asia Pacific Center for Theoretical Physics, Pohang, Gyeongbuk 37673, Korea}

\date{\today}

\begin{abstract}
In-medium valence-quark distributions of $\pi^+$ and $K^+$ mesons in symmetric nuclear matter are studied 
by combining the Nambu--Jona-Lasinio model and the quark-meson coupling model.
The in-medium properties of the current quarks, which are used as inputs for studying the 
in-medium pion and kaon properties in the Nambu--Jona-Lasinio model, 
are calculated within the quark-meson coupling model. 
The light-quark condensates, light-quark dynamical masses, pion and kaon decay constants, 
and pion- and kaon-quark coupling constants are found to decrease as nuclear density increases. 
The obtained valence quark distributions in vacuum for both the $\pi^+$ and $K^+$ could reasonably 
describe the available experimental data over a wide range of Bjorken-$x$. 
The in-medium valence $u$-quark distribution in the $\pi^+$ at $Q^2=16~\mbox{GeV}^2$ is found to be almost 
unchanged compared to the in-vacuum case. 
However, the in-medium to in-vacuum ratios of both the valence $u$-quark and valence $s$-quark distributions of the $K^+$ meson
at $Q^2=16~\mbox{GeV}^2$ increase with nuclear matter density, but show different $x$-dependence.
Namely, the ratio for the valence $u$-quark distribution increases with $x$, while that for the valence $s$ quark 
decreases with $x$.
These features are enhanced at higher density regions.
\end{abstract}

\maketitle

\section{Introduction} \label{intro}
%
Pions and kaons, which emerge as a consequence of dynamical chiral symmetry breaking~\cite{NJL61a}, 
play very important and special roles in understanding the nonperturbative features of low-energy 
Quantum Chromodynamics (QCD)~\cite{Ioffe05}. 
The nonperturbative aspects of QCD can be accessed through the elastic electromagnetic form factors 
and quark distribution functions of these mesons~\cite{GPV01,BR05,HCT16} as well as hadron distribution functions in 
the low-$Q^2$ region, where the latter quantities are used as inputs for $Q^2$-evolution to high-$Q^2$ region 
that the perturbative QCD phenomena are experimentally accessible. 
They can provide us with important information on the internal structure of pions and kaons as well as 
the quark and gluon dynamics. 
However, it is highly nontrivial to calculate the quark distribution functions of these mesons in vacuum at low-energy scale. 
Several models~\cite{HCT16,HBCT18,RA08,NK07b,BT02,Zovko74,DDEF12,BM88a,Hutauruk18,CCRST13,FJ79}, and 
lattice QCD simulations~\cite{HPQCD-17,OKLMM15,HKLLWZ07,PACS-17} 
have been applied to study the internal structure of pions and kaons in vacuum. 
Despite of these efforts, theoretical understanding of quark distributions of pions and kaons requires more efforts and 
the situation is worsened by the scarce of experimental data~\cite{NA7-86,ABBB84,JLAB_Fpi-08}, in particular, for kaons.
Attempts to understand quark distribution functions of mesons in a nuclear medium are much more restricted.

However, some recent progress has been reported for understanding the quark distributions of pions and kaons. 
Several studies were made to uncover the internal structure of pions in a nuclear medium, 
e.g., pion properties in nuclei~\cite{MSTV90,DTERF14}, and pion photoproduction off nuclei~\cite{BL77}.
Furthermore, for the strange quark sector, investigations on kaon photoproduction off nuclei~\cite{LMBW99} and 
kaon-nucleus Drell-Yan processes~\cite{LLT96} were reported. 
The first evidence that triggered the interests for hadron properties in a nuclear medium is the EMC effects reported by
the European Muon Collaboration (EMC) in Ref.~\cite{EMC83b}, where the deep inelastic scattering data revealed that parton distributions 
of bound nucleons in nuclei are modified and quite different from those of free nucleons~\cite{GST95}.
It is, therefore, natural to expect that the internal structure of mesons would also be modified in a nuclear medium 
and affect the meson production in heavy ion collisions (HIC)~\cite{NA49-07}, the reactions involving finite nuclei, 
and the quark distributions of mesons.
A deeper understanding of the in-medium modifications of the quark distributions 
of mesons can provide us with useful information on the hadronization phenomena 
in HIC~\cite{NA49-07}, pion- and kaon-nuclear Drell-Yan processes~\cite{LLT96,Saclay-80}, 
as well as the recently proposed experimental measurements of structure  
functions of pions and kaons via the pion- and kaon-induced Drell-Yan 
production on polarised and unpolarised proton, deuteron, and nuclear targets 
at the $M2$ beam line of CERN SPS~\cite{COMPASS18}.

Recently, the electromagnetic form factors of pions and kaons~\cite{DTERF14,YDDTF17} as well as 
meson distribution amplitudes~\cite{TD16} in a nuclear medium were studied by combining the light-front constituent 
quark model and the quark-meson coupling (QMC) model.
In these publications, the QMC model was used to provide the in-medium light-quark properties, which were then used as inputs 
for studying the medium modifications of meson properties. 
However, quark distributions, or parton distribution functions (PDFs) of mesons in a nuclear medium should be 
explored further in more elaborated manner.   
For example, consistency with the chiral limit of QCD is an important constraint for a model study, which is not observed
by the previous studies.
This is one of the main motivations of the present work.
An early attempt to investigate the pion structure in a nuclear medium was made around two decades ago, by 
combining the Nambu--Jona-Lasinio (NJL) model and the operator product expansion~\cite{Suzuki95} to incorporate 
the nuclear medium effects in the calculations of twist-2 operators. 
It was found that the in-medium structure function of the pion decreases in the large-Bjorken-$x$ region 
at the saturation density $\rho_0^{} =0.17~\mbox{fm}^{-3}$.
In the present work we study the valence-quark distributions (or valence PDFs) of both the $\pi^+$ and $K^+$ mesons 
in symmetric nuclear matter based on the NJL model
with an improved treatment for the nuclear medium effects. 
Namely, the in-medium quantities, which are served as inputs in the present study, are consistently constrained by the symmetric 
nuclear matter properties at saturation density in the QMC model, and the NJL model is used to satisfy the chiral limit.

One of the main focuses of the present work is on the in-medium modifications of the valence $u$-quark distributions in 
$\pi^{+}$ and $K^{+}$ mesons in symmetric nuclear matter. 
Our approach is based on the NJL model with the constrained in-medium inputs calculated by the QMC model.
The NJL model is a powerful chiral effective quark theory of QCD with numerous successes in the studies of 
meson properties~\cite{BJM88,BM88,VW91,Klevansky92,CBT14}.
Recently, the NJL model was applied for studying the kaon electromagnetic form factor and valence-quark distributions 
in vacuum~\cite{HCT16}, as well as the in-medium electroweak properties of pions~\cite{HOT18b}.
By extending the works of Refs.~\cite{HCT16,HOT18b}, we explore the valence-quark distributions of the $\pi^+$ and $K^+$ mesons,  
dynamical quark masses, and pion/kaon properties in symmetric nuclear matter.

This paper is organized as follows. 
In Sec.~\ref{vacnjlprop} we briefly review the calculation of the meson properties in the NJL model.
The in-medium quark properties in the QMC model are described in Sec.~\ref{QMCprop},
which will be used as inputs to study the in-medium pion and kaon properties based on the NJL model in Sec.~\ref{pionNJLmedium}. 
Presented in Sec.~\ref{mediumstructurefunction} are the in-medium modifications of valence-parton distribution 
functions of $\pi^+$ and $K^+$ mesons calculated in the present work. 
The implications of our numerical results are also discussed.
Section~\ref{summary} gives the summary.

\section{Pion and Kaon Free-Space Properties in the NJL model} \label{vacnjlprop}

In this section we briefly review how we calculate the pion and kaon free-space properties in the NJL model. 
(See Refs.~\cite{CBT14,HCT16} for details.)
The NJL model is an effective theory of QCD in the low-energy region, which contains important features of QCD 
such as chiral symmetry and its breaking pattern. 
Thus it has been widely used for understanding various low-energy nonperturbative phenomena of QCD~\cite{Klevansky92}.

To study kaon properties, one needs to extend the NJL model to three-flavor case of which Lagrangian with four-fermion 
contact interactions reads 
\begin{eqnarray}
\label{eq:lagNJL}
\mathscr{L}_{\rm NJL} & = & \bar{\psi} ( i \slashed{\partial} - \hat{m} ) \psi 
+ G_\pi \left[ (\bar{\psi} \lambda_a^{} \psi )^2 - (\bar{\psi} \lambda_a^{} 
\gamma_5^{} \psi )^2 \right]
\nonumber \\ &&
- G_\rho \left[ (\bar{\psi} \lambda_a^{} \gamma^\mu \psi )^2
  + (\bar{\psi} \lambda_a^{} \gamma^\mu \gamma_5^{} \psi )^2 \right] ,
\end{eqnarray}
where a sum over $a=(0, \dots, 8)$ is understood and $\lambda_a$ are the Gell-Mann matrices in flavor space with 
$\lambda_0 \equiv \sqrt{\frac{2}{3}} \mathbbm{1}$.
The quark field $\psi$ represents $\psi = (u, d, s)^T$ and 
$\hat{m} = \mbox{diag} (m_u, m_d, m_s)$ is 
the current-quark mass matrix. 
The four-fermion coupling constants are represented by $G_\pi$ and $G_\rho$.%
\footnote{In principle, the two pieces in the $G_\rho$ term of Eq.~\eqref{eq:lagNJL} may have different coupling constants 
since they are individually chiral invariant. 
Our choice of the common coupling avoids flavor mixing and gives the flavor content of $\omega$-meson as 
$(u \bar{u} + d \bar{d})/\sqrt2$ and $\phi$-meson as $s\bar{s}$, that is, we assume the ``ideal mixing'' for these mesons.}
Throughout this work, we assume $m_l \equiv m_u = m_d$, where $l=( u, d)$ 
stands for ``light'', and thus $m_l$ is the light-quark mass.

The general solution to the standard NJL gap equation has the form of
\begin{equation}
  S^{-1}_q(k) = \slashed{k} - M_q + i \epsilon,
\end{equation}
for a quark flavor $q$ ($= u, d, s$) and the dynamically generated constituent quark mass $M_q$ is obtained as
\begin{eqnarray}
\label{eq:masNJL}
M_q &=& m_q - 4 \, G_\pi \braket{ \bar{q} q }
\nonumber \\
&=& m_q^{} + 12 i G_\pi \int \frac{d^4 k}{(2\pi)^4} \mbox{Tr}_D^{} [S_q (k)].
\end{eqnarray}
Here, the quark condensate is denoted by $\braket{ \bar{q} q }$ and the trace is only for the Dirac-space indices.
Then the proper-time regularization scheme leads to
\begin{align}
  \label{eq:massNJLProp}
  M_q &= m_q^{} + \frac{3G_\pi M_q}{\pi^2}
  \int_{1/\Lambda_{\rm UV}^{2}}^{1/\Lambda_{\rm IR}^{2}} \frac{d\tau}{\tau^2} 
\exp\left(-\tau M_q^2\right),
\end{align}
where $\Lambda_{\rm UV}^{}$ and $\Lambda_{\rm IR}^{}$ are, respectively, the ultraviolet and infrared cutoff parameters.

Pions and kaons are described in the NJL model as quark-antiquark bound states by solving the corresponding Bethe-Saltpeter equations (BSEs). 
The solutions to the BSEs are given by a two-body $t$-matrix that depends on the interaction channel. 
For the pseudoscalar channel $\alpha$ ($= \pi, K$), 
it is given by 
\begin{align}
\label{eq:tmatrix}
\tau_\alpha(p) = \frac{-2i\,G_\pi}{1 + 2\,G_\pi\,\Pi_\alpha (p^2)}, 
\end{align}
where the bubble diagrams lead to 
\begin{align}
& \Pi_{\pi}(p^2) = 6i \int \frac{d^4k}{(2\pi)^4}\ \mbox{Tr}_D^{} \left[ \gamma_5^{} \,S_{l}(k) \gamma_5^{} \,S_{l} (k+p) \right], \\
& \Pi_{K}(p^2) = 6i \int \frac{d^4k}{(2\pi)^4}\ \mbox{Tr}_D^{} \left[ \gamma_5^{} \,S_{l}(k) \gamma_5^{} \,
S_{s} (k+p) \right].
\end{align}

The meson masses are identified by the pole positions in the corresponding $t$-matrices. 
Thus, the pion and kaon masses are given, respectively, 
by the solutions of the equations, 
\begin{align}
  1 + 2\, G_\pi\, \Pi_{\pi} (p^2 = m_{\pi}^2) &= 0, \nonumber \\
  1 + 2\, G_\pi\, \Pi_{K} (p^2 = m_{K}^2) &= 0.
\end{align}  
These equations can be rearranged to give the pion and kaon masses as
\begin{align}
  m_{\pi}^2 &= \frac{m_l}{M_l} \frac{2}{G_\pi\, \mathcal{I}_{ll}(m_\pi^2)}, \nonumber \\
  \label{eqn:mesonmass}
  m_{K}^2 &= \left(\frac{m_s}{M_s} + \frac{m_l}{M_l} \right)
\frac{1}{G_\pi \mathcal{I}_{ls} (m_K^2)} + (M_s -M_l)^2,  
\end{align} 
where
\begin{align}
\mathcal{I}_{ab}(p^2) &= \frac{3}{\pi^2} 
\int_0^1 dz \int_{1/\Lambda^2_{\rm UV}}^{1/\Lambda^2_{\rm IR}} \frac{d\tau}{\tau}\,
e^{-\tau[-z(1-z)\,p^2 + z\,M_b^2 + (1-z)\,M_a^2]}\, , 
\end{align} 
for quark flavors $a$ and $b$.
Equation~(\ref{eqn:mesonmass}) makes it evident that chiral symmetry and its breaking pattern are embedded in the NJL model. 
The model satisfies the chiral limit and the pions and kaons become massless in the chiral limit being realized as Goldstone bosons. 
The residue at a pole in the $\bar{q}q$ $t$-matrix defines the meson ($\alpha$)-quark coupling constant $g_{\alpha q q}$ as
\begin{align}
\label{eq:couplinconstant}
g_{\alpha q q}^{-2} &= - \left.\frac{\partial\, \Pi_\alpha (p^2)}{\partial p^2} 
\right|_{p^2 = m_\alpha^2} 
\end{align}
with $\alpha = \pi, K$.

The pion and kaon weak decay constants are determined from the meson to vacuum transition matrix element 
$\braket{ 0 | \mathscr{J}_a^{5 \mu} (0) | \alpha (p) }$ with $\mathscr{J}_a^{5 \mu}$ being the quark weak axial-vector current 
operator for a flavor quantum number $a$.
Evaluating the corresponding matrix elements, pion and kaon weak decay constants in the proper-time regularization
scheme are expressed as~\cite{Hutauruk16,NBC14}
\begin{align}
  \label{eq:decayconNJL}
f_\pi &= \frac{3 g_{\pi qq}^{} M_q}{4 \pi^2} \int_0^1 dz
\int_{1/\Lambda_{\rm UV}^{2}}^{1/\Lambda_{\rm IR}^{2}} 
\frac{d\tau}{\tau} e^{-\tau [ M_q^2 - z(1-z) m_\pi^2 ]}, \nonumber \\
f_K &= \frac{3 g_{K qq}^{}}{4 \pi^2} \int_0^1 dz
\int_{1/\Lambda_{\rm UV}^{2}}^{1/\Lambda_{\rm IR}^{2}} \frac{d\tau}{\tau} 
\left[ M_s + z(M_l - M_s) \right] \nonumber \\
	&\hspace{18ex} \times e^{-\tau [ M_s^2 - z(M_s^2 -M_l^2)-z(1-z) m_K^2]}.
\end{align}
This completes the formalism for meson masses, decay constants, and meson-quark coupling constants in vacuum.

Following Refs.~\cite{CBT14,HCT16}, we choose $\Lambda_{\rm IR} = 240$~MeV which is 
set based on $\Lambda_{\rm QCD}$ associated with confinement, and
$M_l = 400$~MeV. 
The model parameters are then fixed to give $m_\pi^{} = 140$~MeV and $m_K^{} = 495$~MeV, together
with the pion decay constant $f_\pi = 93$~MeV. 
This procedure gives $\Lambda_{\rm UV} = 645$~MeV, $G_\pi = 19.0~\mbox{GeV}^{-2}$, and $M_s = 611$~MeV.
The corresponding current quark masses are $m_{l} = 16.4$~MeV and $m_s=  356$~MeV.
The predicted values for the kaon decay constant and $u$ and $s$ quark condensates in vacuum are 
$f_{K}=91$~MeV, $\braket{\bar{u} u}^{1/3}= -171$~MeV, and $\braket{ \bar{s} s }^{1/3}= -151$~MeV, respectively.

\section{In-medium quark properties in the QMC model} \label{QMCprop}

In the present approach, the in-medium current-quark properties are provided by the QMC model~\cite{Guichon88},  
which will be used in the NJL model to explore the in-medium dynamical quark and meson properties. 
The QMC model has been successfully applied for studying many topics in nuclear and hadron physics such as finite 
nuclei~\cite{GSRT95,STT96b,STT96,SGRT16,GST18}, hypernuclei~\cite{TSHT98,GTT07}, superheavy nuclei~\cite{SGT17}, 
neutron star properties~\cite{WCTTS13}, and nucleon/hadron properties in a nuclear medium~\cite{STT07,KTT17}.
In the QMC model, medium effects are incorporated by the self-consistent exchanges of the scalar ($\sigma$), vector-isoscalar ($\omega$), and 
vector-isovector ($\rho$) meson fields, which directly couple to the confined light quarks inside the nucleon, rather than to a pointlike nucleon. 
In the following, we consider symmetric nuclear matter in its rest frame in 
the Hartree mean-field approximation. 
(See Ref.~\cite{KTT98} for detailed discussions on the Hartree-Fock treatment.)

The effective Lagrangian for symmetric nuclear matter is given by~\cite{GTT07,STT07}
\begin{align}
  \label{eqintro1}
  \mathscr{L}_{\textrm{QMC}} & =  \bar{\psi}_N \left[ i\gamma \cdot 
\partial - M_{N}^{*}(\sigma)
    - g_\omega \omega^{\mu} \gamma_\mu \right] \psi_N + \mathscr{L}_\textrm{meson},
\end{align}
where $\psi_N$, $\sigma$, and $\omega$ are the nucleon, $\sigma$-, and $\omega$-meson fields, respectively.
The effective nucleon mass $M_N^{*}$ is defined by
\begin{align}
  M_{N}^{*} \left( \sigma \right) &= M_{N} - g_\sigma \left( \sigma \right) \sigma.
\end{align}
Here, $g_{\sigma} ( \sigma )$ and $g_{\omega}$ are the $\sigma$-dependent nucleon-$\sigma$ and
nucleon-$\omega$ coupling constants, respectively.
We define the nucleon-$\sigma$ coupling constant as $g^N_\sigma \equiv g_\sigma (\sigma=0)$ for later convenience.
Because symmetric nuclear matter is isospin saturated, the isospin-dependent $\rho$-meson field vanishes in the Hartree approximation
and is not included in Eq.~(\ref{eqintro1}).
The free meson Lagrangian density in Eq.~(\ref{eqintro1}) is given by 
\begin{align}
  \label{eqintro3}
  \mathscr{L}_\textrm{meson} &= \frac{1}{2} (\partial_\mu \sigma
  \partial^\mu \sigma - m_\sigma^2 \sigma^2)
  - \frac{1}{2} \partial_\mu \omega_\nu (\partial^\mu \omega^\nu
  - \partial^\nu \omega^\mu)\nonumber \\
  &+ \frac{1}{2} m_\omega^2 \omega^\mu \omega_\mu.  
\end{align}

In the Hartree mean-field approximation the nucleon Fermi momentum $k_F^{}$ is related to the baryon density ($\rho_{B}^{}$)
and scalar density ($\rho_{s}^{}$) defined as
\begin{align}
  \label{eqintro4}
  \rho_{B}^{} &= \frac{\gamma}{(2\pi)^3} \int d\bm{k}\, \Theta \left( k_F^{}  - | \bm{k} | \right) =
  \frac{\gamma k_F^3}{3 \pi^2}, \nonumber \\
  \rho_{s}^{} &= \frac{\gamma}{(2\pi)^3} \int d\bm{k}\,  \Theta \left( k_F^{} - | \bm{k} | \right)
  \frac{M_N^{*} (\sigma)}{\sqrt{M_N^{*2} (\sigma ) + \bm{k}^2}},
\end{align}
where $\Theta(x)$ is the Heaviside step function and $\gamma =4$ for symmetric nuclear matter. 
The baryon density $\rho_B$ is given by $\rho_B^{} = \rho_p + \rho_n$, where $\rho_p$ and $\rho_n$ are proton and neutron densities,
respectively.

In the QMC model, nuclear matter is described as a collection of nonoverlapping MIT bags of nucleons~\cite{CJJTW74}.
The Dirac equations for the quarks and antiquarks in the bag are given by
\begin{align}
  \label{eqintro5}
  \left[ i \gamma \cdot \partial_{x} - \left( m_l^{} - V_{\sigma}^{q} \right) 
\mp \gamma^{0} \left( V_{\omega}^{q}
    + \frac{1}{2} V_{\rho}^{q} \right) \right] \left( \begin{array}{c}\psi_u(x)  \\ 
\psi_{\bar{u}}(x) \\ \end{array} \right) &= 0, \nonumber \\
  \left[ i \gamma \cdot \partial_{x} - \left( m_l^{} - V_{\sigma}^{q} \right) 
\mp \gamma^{0} \left( V_{\omega}^{q}
    - \frac{1}{2} V_{\rho}^{q} \right) \right] \left( \begin{array}{c} \psi_d(x)  \\ 
\psi_{\bar{d}}(x) \\ \end{array} \right) &= 0, \nonumber \\
  \left[ i \gamma \cdot \partial_{x} - m_{s} \right] \left( \begin{array}{c}  \psi_s(x)  \\ 
\psi_{\bar{s}}(x) \\ \end{array} \right) &= 0,
\end{align}
and we can define the effective in-medium current quark mass $m_l^{*}$ as
\begin{align}
  \label{eqintro5a}
m_l^{*} & \equiv m_l^{} - V_{\sigma}^{q},
\end{align}
with $-V_\sigma^{q}$ being the scalar potential, while $m_s^* = m_s^{}$.
The scalar and vector mean-field potentials felt by the light quarks in symmetric nuclear matter are, respectively, defined by
\begin{align}
  V_{\sigma}^{q} & \equiv g_{\sigma}^{q} \sigma = g_{\sigma}^{q} \braket{\sigma},
\nonumber \\
V_{\omega}^{q} &\equiv g_{\omega}^{q} \omega = g_{\omega}^{q} \, \delta^{\mu , 0} \braket{ \omega^{\mu} },
\end{align}
where $g_{\sigma}^{q}$ and $g_{\omega}^{q}$ are the coupling constants of the light quarks
to the mean-field $\sigma$ and $\omega$, respectively. Note that in the present approach
the strange quark is decoupled from the scalar and vector mean-field potentials in nuclear medium.

The bag radius of hadron $h$ in a nuclear medium, $R_h^{*}$, is determined through the stability condition of the hadron mass. 
The eigenenergies of the quarks, in units of $1/ R_h^{*}$, in the MIT bags are obtained as
\begin{align}
  \label{eq:kaonmed9}
  \left( \begin{array}{c}
    \epsilon_u \\ \epsilon_{\bar{u}}
  \end{array} \right)
  & = \Omega_l^* \pm R_h^* \left( V^q_\omega + \frac{1}{2} V^q_\rho \right), \nonumber \\
  \left( \begin{array}{c} \epsilon_d \\  \epsilon_{\bar{d}}
  \end{array} \right)
  &= \Omega_l^* \pm R_h^* \left( V^q_\omega - \frac{1}{2} V^q_\rho \right), \nonumber \\
    \left( \begin{array}{c} \epsilon_s \\  \epsilon_{\bar{s}}
  \end{array} \right)
  & = \Omega_{s}.
\end{align}
The effective mass of hadron $h$, $ m_h^{*}$, in nuclear medium is calculated by 
\begin{align}
  \label{eq:kaonmed10}
  m_h^{*} &= \sum_{j = l, \bar{l}, s, \bar{s}} \frac{n_j \Omega_j^{*} -z_h}{R^{*}_h}
  + \frac{4}{3} \pi R_h^{* 3} B,
\end{align}
which determines $R_h^{*}$ by the stability condition, i.e.,
\begin{align}
  \left. \frac{d m_h^{*}}{d R_h} \right \vert_{R_h = R_h^{*}} = 0,
\end{align}
where $\Omega^{*}_l = \Omega^{*}_{\bar{l}} = \left [x_l^2 + \left(R_h^{*} m_l^{*} \right)^2 
\right]^{1/2}$ and $\Omega_{\bar{s}}^{*} = \Omega_{s}^{*} = [x_{s}^2 
+ (R_h^{*} m_s )^2 ]^{1/2}$.
The parameter $z_h$ accounts for the sum of the center-of-mass and gluon fluctuation corrections and is assumed to be independent 
of density~\cite{SGRT16}, and $B$ is the bag constant. 
For the quark in the bag of hadron $h$, the ground state wave function satisfies the boundary condition at the bag surface, 
$j_0 (x) =  \beta_q j_1 (x)$, where
\begin{align}\
  \label{eq:beta}
  \beta_q &= \sqrt{\frac{\Omega_q^{*} -m_q^{*} R_h^{*}}{\Omega_q^{*} + m_q^{*} R_h^{*}}},
\end{align}
and $j_{0,1}$ are spherical Bessel functions.
This determines the values of $x_l$ and $x_s$.

Except for the saturation density and the binding energy at the saturation point that are used to fix the quark-meson coupling constants, 
nuclear matter properties depend on the light-quark current mass in vacuum. 
In Table~\ref{tab:model1} we list two sets of the QMC model results corresponding to two different current-quark mass values 
as well as the corresponding calculated quantities.
The first row shows the quantities obtained with the standard QMC model values, $m_l = 5~\mbox{MeV}$ and $m_s = 250~\mbox{MeV}$.
In the present work, however, since we use the NJL model, we adopt $m_l =  16.4~\mbox{MeV}$ 
and $m_s = 356~\mbox{MeV}$ following Ref.~\cite{HCT16}.
The bag radius of the nucleon in free space, $R_N = 0.8~\mbox{fm}$, is also used as an input.


\begin{table}[t]
  \caption{Current-quark masses of light quarks in vacuum and the corresponding  
coupling constants, bag constant $B$, the parameter $z_N$, effective nucleon mass $M_N^{*}$, and the nuclear incompressibility $K$, 
at saturation density, $\rho_0^{} = 0.15~\mbox{fm}^{-3}$ obtained in the QMC model.   
The units of $m_l$, $M_N^*$, and $B^{1/4}$ are MeV.
The standard QMC model quark masses are $m_l = 5$~MeV and $m_s = 250$~MeV, while, in this work, 
we use $m_l = 6.4$~MeV and $m_s = 356$~MeV to be consistent with the NJL model calculations.
}

\label{tab:model1}
\addtolength{\tabcolsep}{2.pt}
\begin{tabular}{ccccccc} 
\hline \hline
	$m_l$ & $(g^N_{\sigma})^{\, 2}  / 4 \pi$ & $g_{\omega}^2 / 4 \pi$ 
& $B^{1/4}$ & $z_N$ & $M_N^{*}$ & $K$ \\[0.2em] 
\hline
$5 $   & $5.393$ & $5.304$ & $170.0$ & $3.295$ & $754.55$ & $279.30$ \\ 
$16.4$   & $5.438$ & $5.412$ & $169.2$ & $3.334$ & $751.95$ & $281.50$
\\ \hline \hline
\end{tabular}
\end{table}

The scalar and vector meson mean fields at the hadron level can be related with the baryon and scalar  densities by 
\begin{align}
  \label{eq:kaonmed11}
  \omega &= \frac{g_\omega \rho_B^{} }{m_\omega^2}, \nonumber \\
  \sigma &= \frac{4 g^N_\sigma C_N (\sigma)}{(2\pi)^3m_{\sigma}^2}
\int d\bm{k}\, \Theta (k_F - | \bm{k} | ) 
\frac{M_N^{*} (\sigma)}{\sqrt{M_N^{*2} (\sigma) + \bm{k}^2}}, \nonumber \\
	&=  \frac{4 g^N_\sigma C_N (\sigma)}{(2\pi)^3m_{\sigma}^2} \rho_s,
\end{align}
where 
\begin{align}
  C_N (\sigma) &= \frac{-1}{g^N_\sigma} 
\left[ \frac{\partial M_N^{*} (\sigma )}{\partial \sigma } \right],
\end{align}
which yields $C_N (\sigma)=1$ for a pointlike nucleon~\cite{SW86,SW97}.
This is the origin of the novel saturation properties in the QMC model and contains the quark dynamics of nucleons (hadrons). 
Solving the self-consistent equation for the scalar mean field of Eq.~(\ref{eq:kaonmed11}), 
the total energy per nucleon is calculated as
\begin{align}
  \label{eq:kaonmed12}
  E^{\rm tot}/A &= \frac{4}{(2\pi)^3 \rho_B^{}}  \int d\bm{k}\, \Theta (k_F - | \bm{k} |) \sqrt{M_N^{*2} (\sigma) + \bm{k}^2} 
  \nonumber \\
  &+ \frac{m_\sigma^2 \sigma^2}{2\rho_B^{}} + \frac{g_\omega^2 \rho_B^{}}{2 m_\omega^2}.
\end{align}
The quark level coupling constants $g_\sigma^q$ and $g_\omega^q$ are fitted by the binding energy of 15.7~MeV of symmetric nuclear
matter at the saturation density $\rho_0$ with $g^N_\sigma = 3 g^q_\sigma S_N (\sigma= 0)$  and $g_\omega = 3g_\omega^q$, 
where $ g_\sigma^q \simeq 5.6251$ for $m_l = 16.4$~MeV and $S_N(\sigma)$ is given by~\cite{KTT17}
\begin{eqnarray}
\dfrac{\partial M_{N}^*(\sigma)}{\partial \sigma} 
	&=& - 3 g_{\sigma}^q \int_{\rm bag} d^3y \bm{} 
\ {\overline \psi}_q(\bm{y})~\psi_q(\bm{y})
\nonumber \\
&\equiv& - 3 g_{\sigma}^q S_{N}(\sigma) = - \dfrac{\partial}{\partial \sigma}
\left[ g_\sigma(\sigma) \sigma \right],
\label{Ssigma}
\end{eqnarray}
which gives $S_N(\sigma=0) \approx 0.4899$ for $m_l = 16.4$~MeV.
Here, $\psi_q$ is the lowest mode light-quark bag wave function 
obtained by solving the Dirac equation self-consistently in the scalar-$\sigma$ 
and vector-$\omega$ mean fields.

\begin{figure}[t]
  \centering\includegraphics[width=0.95\columnwidth]{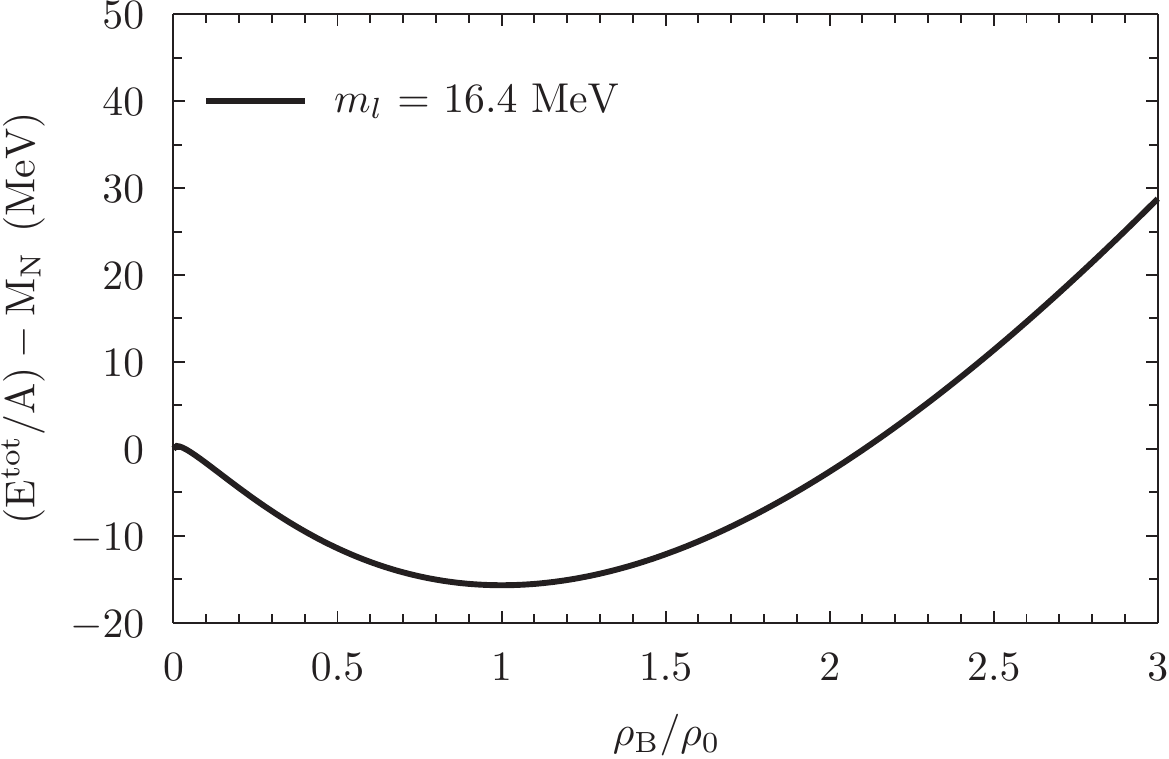}
  \caption{\label{fig1} 
Negative of binding energy per nucleon ($E^{\rm tot} /A - M_N $) for symmetric nuclear matter 
obtained in the QMC model with the free-space quark mass $m_l = 16.4~\mbox{MeV}$.}
\end{figure}

\begin{figure}[t]
  \centering\includegraphics[width=0.95\columnwidth]{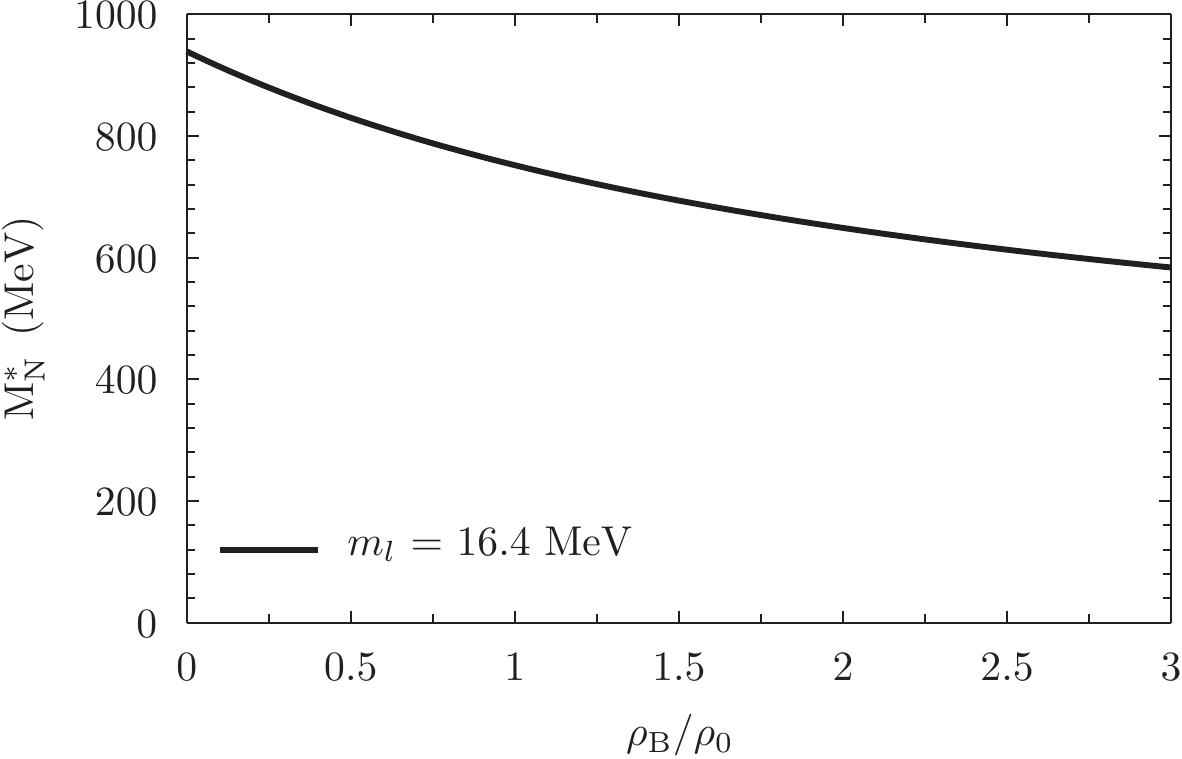}
  \caption{\label{fig2} 
Effective nucleon mass $M^{*}_N$ for symmetric nuclear matter calculated by the 
QMC model with $m_l = 16.4~\mbox{MeV}$.}
\end{figure}

\begin{figure}[t]
\centering\includegraphics[width=0.95\columnwidth]{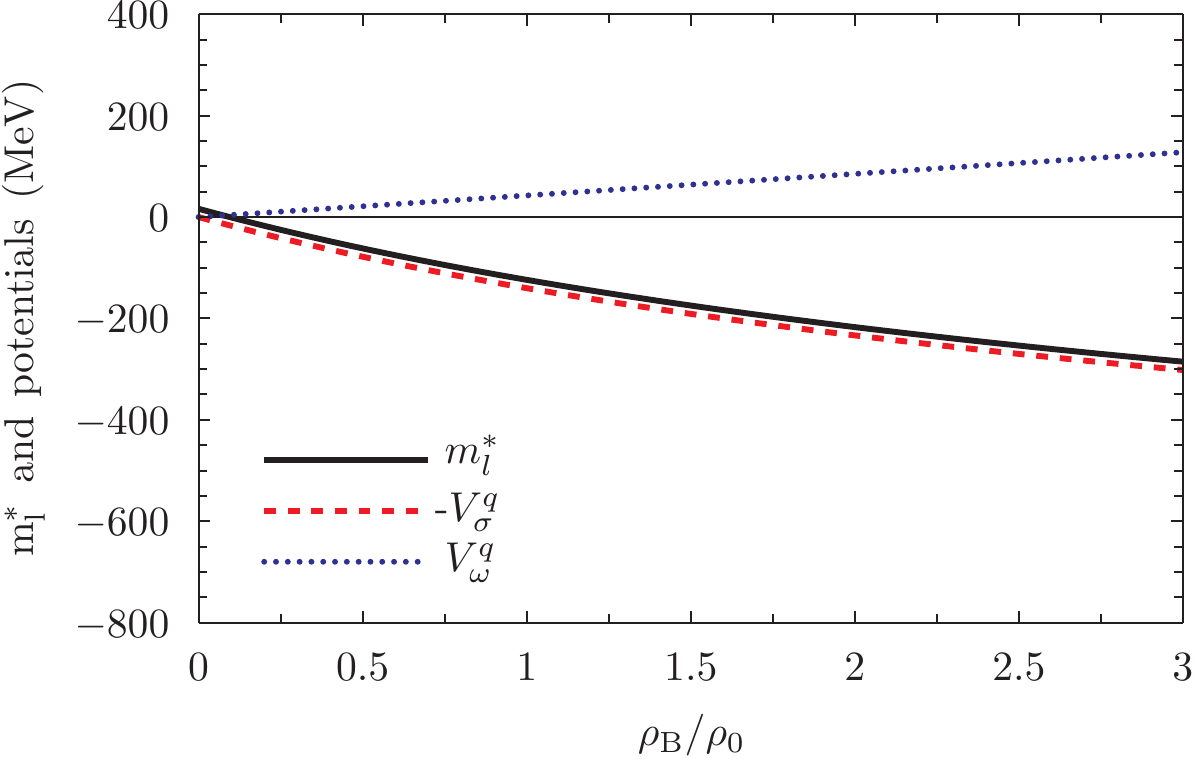}
\caption{\label{fig3} 
Effective current quark mass $m^*_l $ of light quarks (solid line), scalar mean-field  potential $-V^q_\sigma$ (dashed line), 
and vector mean-field potential $V_\omega^q$ (dotted line) calculated by the QMC model with $m_l = 16.4~\mbox{MeV}$.}
\end{figure}

Shown in Fig.~\ref{fig1} is the obtained negative of binding energy per nucleon for symmetric nuclear matter. 
The calculated  effective nucleon mass $M_N^{*}$ is illustrated in Fig.~\ref{fig2} as a function of baryon density.
The corresponding effective light-quark current mass $m_l^*$, scalar mean-field potential $- V^q_\sigma$, 
and vector mean-field potential $V_\omega^q$ are also presented in Fig.~\ref{fig3}.

\section{In-Medium Pion and Kaon Properties} \label{pionNJLmedium}

Based on the combined approach of the NJL-model formalism equipped with in-medium quark properties in the QMC model,
we now explore the in-medium pion and kaon properties in this section. 
The in-medium mean-field potentials felt by the light quarks are computed in the QMC model and are shown in Fig.~\ref{fig3}.
These in-medium properties of current quarks are used to estimate the in-medium dynamical quark masses in the NJL model,  
and they allow us to study the pion and kaon properties in medium such as the medium modifications of the valence-quark distributions 
of pions and kaons. 


The gap equation for the dynamical quark mass $M^*_q$ for a quark $q$ in medium can be 
straightforwardly read from Eq.~(\ref{eq:masNJL}), 
\begin{align}
  \label{eq:kaonmed13}
  M_q^{*} & = m_q^{*} + \frac{3G_\pi M_q^{*}}{\pi^2}
  \int_{1/\Lambda_{\rm UV}^2}^{\infty} \frac{d\tau}{\tau^2} e^{-\tau M_q^{*2}},
\end{align}
where $m_q^{*}$ and $M_q^{*}$ are, respectively, the corresponding density-dependent in-medium current- and dynamical-quark 
masses. 
We use the free space values for the coupling constant $G_\pi$ and the ultraviolet cutoff $\Lambda_{\rm UV}$, while 
we use $\Lambda_{\rm IR} \simeq 0$ or $(1/\Lambda_{\rm IR}^2) = \infty$, since we do not have information on  
$\Lambda_{\rm IR}$ (or $\Lambda_{\rm QCD}$) in medium and the results are not affected much by the value of $\Lambda_{\rm IR}$ when
$1/\Lambda_{\rm IR}^2 \to \infty$.
In the present approach, the above relation holds only for light quarks, whille the strange quark mass does not change in nuclear matter.
In addition, the extra density dependent term introduced in Refs.~\cite{Maedan09,WMT15}, which is proportional to quark chemical potentials 
to treat finite density systems, is not included because the information of symmetric nuclear matter saturation, and thus the medium effects, 
are self-consistently included in the in-medium light-quark properties calculated in the QMC model.

The in-medium dressed quark propagators are expressed by  
\begin{align}
  \label{eq:kaonmed14}
  S_l^{*}(k^*) &= \frac{\slashed{k}^* + M_l^{*}}{(k^*)^2 - (M_l^{*})^2 
  + i \epsilon},
\\
S_s^{*}(k^*) &= S_s(k) =\frac{\slashed{k} + M_s}{k^2 - M_s^2 + i \epsilon},
\end{align}
where the in-medium modifications enter as the shift of the light-quark momenta through $k^{\mu} \to k^{*\mu}=k^\mu + V^\mu$.
Here, $V^\mu = (V^0, \textbf{0})$ is the vector mean-field potential defined in the previous section~\cite{STTS98,STT04,CMPR09}.
An asterisk over a quantity denotes an in-medium quantity as before.

As in the vacuum case, mesons are described as dressed quark-antiquark bound states that appear 
as solutions of the BSE in the random phase approximation.
The solution to the BSE in each meson channel is given by the $t$-matrix of a two-body scattering that depends 
on the nature of the interaction channel. 
The in-medium reduced $t$-matrices for $\pi$ and $K$ mesons take the same form as in vacuum:
\begin{align}
  \label{eq:tmatrix_pi}
  \tau^{*}_\pi (p^{*}) &= \frac{-2i\,G_\pi}{1 + 2 G_\pi \Pi^{*}_\pi (p^{*2})}, \\
  \label{eq:tmatrix_k}
  \tau^{*}_K (p^{*}) &= \frac{-2i\,G_\pi}{1 + 2 G_\pi \Pi^{*}_K (p^{*2})}, 
\end{align}
where the in-medium bubble diagrams lead to
\begin{align}
\Pi^{*}_{\pi} (p^{*2}) &= 6i \int \frac{d^4k}{(2\pi)^4} \, 
\mbox{Tr}_D^{} \left[ \gamma_5^{} S^{*}_{l}(k^*) \gamma_5^{} S^{*}_{l}(k^*+p^*) \right],\\
  \Pi^{*}_{K} (p^{*2}) &= 6i \int \frac{d^4k}{(2\pi)^4}\, 
\mbox{Tr}_D^{} \left[ \gamma_5^{} S^{*}_{l}(k^*) \gamma_5^{} S^{*}_{s}(k^*+p^*) \right].
\end{align}
These equations show that the medium-modified momentum enters for the momenta of light quarks and the kaon also feels 
the vector potential, which is in contrast to the pion case where the vector potentials for the light quark and light antiquark 
cancel out. 
Thus, although the vector potential effect can be eliminated by the integral variable shift for $\Pi^*_\pi$,
it should be explicitly included in the calculation of $\Pi^*_K$.

The meson masses are defined by the poles in the corresponding $t$-matrices as in the vacuum case, and
Eqs.~(\ref{eq:tmatrix_pi}) and~(\ref{eq:tmatrix_k}) lead to
\begin{align}
  1 + 2 G_\pi \Pi^{*}_{\pi} (p^{*2} = m_{\pi}^{* 2}) &= 0,\\
  1 + 2 G_\pi \Pi^{*}_{K} (p^{*2}= m_{K}^{* 2}) &= 0.
\end{align}  
These relations can be rewritten as
\begin{align}
  m_{\pi}^{* 2} &= \frac{m^{*}_l}{M^{*}_l} \frac{2}{G_\pi \mathcal{I}_{ll}(m_\pi^{*2})},
 \nonumber \\
  m_{K}^{* 2} &= \left[ \frac{m^{*}_s}{M^{*}_s}
    + \frac{m^{*}}{M^{*}_l} \right]  \frac{1}{G_\pi \mathcal{I}_{l s}(m_K^{* 2})}
  + (M^{*}_s-M^{*}_l)^2,
\end{align} 
where
\begin{align}
  \mathcal{I}_{\!ab}(p^{* 2}) &= \frac{3}{\pi^2} \int_0^1 dz \int
  \frac{d\tau}{\tau}\, \nonumber \\
	\times &e^{-\tau \left[ -z(1-z)\,p^{*2} + 2p^{*}V^0 z(1-z) - z(1-z)({V^0})^2 
+ z\,M_b^{* 2} + (1-z)\,M_a^{* 2} \right]}.
\end{align} 
As in the vacuum case, the residue at a pole in the $\bar{q}q$ $t$-matrix defines the in-medium coupling constant $g^{*}_{\alpha q q}$ as
\begin{align}
\label{eq:couplinconstantmed}
g_{\alpha q q}^{* -2} &= - \left.\frac{\partial\, 
\Pi_\alpha (p^{*2})}{\partial p^{*2}} \right|_{p^{*2} = m_\alpha^{*2}}.
\end{align}

\begin{table*}[t]
  \caption{Results for the in-medium properties of dynamical quarks and mesons 
in the NJL model calculated with the in-medium quark masses $m^*_l$ and $m^*_s = m_s$ for 
the vacuum values $m_{l} = 16.4$~MeV and  $m_{s} = 356$~MeV. 
The quantities are in units of GeV, except for the meson-quark coupling constant that is dimensionless.
Here, we have $M_s^* = M_s = 0.611$~GeV.}
  \label{tab:model2}
  \addtolength{\tabcolsep}{6.8pt}
  \begin{tabular}{ccccccccc} 
    \hline \hline
    $\rho_B / \rho_0$ & $M_u^{*}$  & $m_K^{*}$ & $f_K^{*}$
    & $g_{K q q}^{*}$ & $m_\pi^{*}$ & $f_\pi^{*}$ & $g_{\pi q q}^{*}$ &
    $-\braket{ \bar{u} u }^{* 1/3}$  \\[0.2em] 
    \hline
    $0.0$  & $0.400$ & $0.495$ & $0.091$ & $4.570$ & $0.140$ & $0.093$ & $4.255$ & $0.171$  \\ 
    $0.25$ & $0.370$ & $0.465$ & $0.091$ & $4.536$ & $0.136$ & $0.092$ & $3.964$ & $0.167$  \\
    $0.50$ & $0.339$ & $0.437$ & $0.090$ & $4.495$ & $0.134$ & $0.089$ & $3.720$ & $0.162$  \\
    $0.75$ & $0.307$ & $0.411$ & $0.089$ & $4.455$ & $0.132$ & $0.086$ & $3.494$ & $0.156$  \\
    $1.00$ & $0.270$ & $0.386$ & $0.088$ & $4.408$ & $0.131$ & $0.081$ & $3.265$ & $0.149$  \\
    $1.25$ & $0.207$ & $0.359$ & $0.084$ & $4.332$ & $0.136$ & $0.069$ & $2.948$ & $0.136$ 
    \\ \hline \hline
  \end{tabular}
\end{table*}

Results for the in-medium properties of dynamical quarks, pions, and kaons
are presented in Table~\ref{tab:model2}.
These results show that the considered in-medium quantities of dynamical quarks and mesons
decrease with increasing density.
At normal nuclear density $\rho_0^{}$, the dynamical $u$-quark mass is found to decrease by about 30\%,  
while the magnitude of the $u$-quark condensate decreases by about 13\%. 
These results indicate that chiral symmetry is partially restored in the finite density system.

In the case of the pion, we find that its mass decreases by about 7\% at normal nuclear density. 
For the pion-quark coupling constant and pion decay constant we find $g_{\pi qq}^*/g_{\pi qq} \approx 0.77$  
and $f_\pi^*/ f_\pi = 0.87$ at normal nuclear density.
Our result for $f_\pi^*/ f_\pi$ is in good agreement with that of Refs.~\cite{KY04,KW97} but is about 10-20\% larger than that  
of Refs.~\cite{MOW01,TW95}.
In the kaon case, the tendency of medium modifications is somehow different as the $s$-quark properties 
are not modified in the present approach because it decouples from mean fields. 
As a result, the medium modifications of kaon properties are realized only through the light quark in the kaon.
Our results show that the kaon mass decreases by about 20\% 
at normal nuclear matter density, 
while both the kaon-quark coupling constant and kaon decay constant, decrease only a few percent.

\section{In-Medium Valence-Quark Distributions of Pions and Kaons} 
\label{mediumstructurefunction}

As quark properties are modified in nuclear medium, it is quite natural to expect that the quark distributions or 
parton distribution functions (PDFs) of mesons are also modified in medium. 
In this section, following the PDF calculations of Ref.~\cite{HCT16}, we evaluate the in-medium valence PDFs 
(or valence-quark distributions) of pions and kaons. 
We start with the twist-2 quark distribution in a hadron $\alpha$ defined by
\begin{align}
  \label{eq:valence1}
  q_\alpha(x) &= p^{+} \int \frac{d\xi^{-}}{2\pi}  e^{ix p^{+} \xi^{-}} 
  \braket{ \alpha | \bar{\psi}_{q}(0) \gamma^{+} \psi_q (\xi^{-}) | \alpha }_{c},
\end{align}
where $c$ denotes the connected-diagram matrix element and $x = {k^+}/{p^+}$ is the Bjorken scaling variable 
with $p^+$ ($k^+$) being the plus-component of the hadron (struck quark) momentum. 
In the NJL model, gluons are ``integrated out'' and the gauge-link, which should appear in Eq.~\eqref{eq:valence1}, is unity.

\begin{figure}[t]
  \centering\includegraphics[width=\columnwidth]{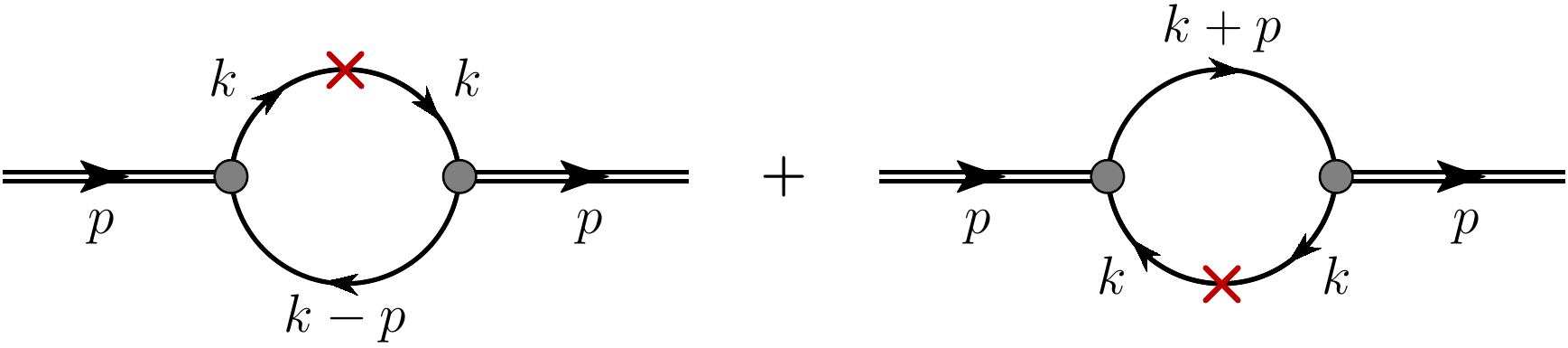}
  \caption{\label{fig:strucfun1}  
Feynman diagrams for the valence-quark distributions in a meson.  
The operator insertion $\gamma^+ \delta \left( p^+x - k^+ \right) \hat{P}_q$, 
where $\hat{P}_q$ is the projection operator for a quark $q$, is represented by the red cross.} 
\end{figure}

The valence-quark distribution functions (or valence PDFs) given by Eq.~(\ref{eq:valence1}) are
calculated based on the two Feynman diagrams depicted in Fig.~\ref{fig:strucfun1}, 
where the operator insertion is given by $\gamma^+\delta \left( p^+x - k^+ \right) \hat{P}_q$, with $\hat{P}_q$ 
being the projection operator for a quark $q$ defined as 
\begin{align}
\hat{P}_{u/d} &= \textstyle \frac{1}{2} \left(  \frac{2}{3}\, \mathbbm{1} \pm \lambda_3 
+ \frac{1}{\sqrt{3}}\,\lambda_8 \right), &
\hat{P}_s &= \textstyle \frac{1}{3}\,\mathbbm{1} - \frac{1}{\sqrt{3}}\,\lambda_8.
\end{align}
Using the relation $\bar{q}(x) = -q(-x)$, the valence-quark and valence-antiquark distributions 
in a meson $\alpha$ are calculated as 
\begin{eqnarray}
\label{eq:valence31}
q_\alpha (x) &=& i g_{\alpha q q}^{*2} \int \frac{d^4k}{(2\pi)^4}\, 
\delta \left( p^{*+}x - k^{*+} \right) 
\nonumber \\ && \mbox{} 
\hspace{-5ex} 
	\times \mathrm{Tr} \left[ \gamma_5\lambda_\alpha^\dagger\,S^{*}(k^*)\,\gamma^+ 
	\hat{P}_q\,S^{*}(k^*)\,\gamma_5\lambda_\alpha\,S^{*}(k^*-p^*) \right], \\
\textbf{\label{eq:valence32}}
\bar{q}_\alpha(x) &=& -i\,g_{\alpha q q}^{*2} \int \frac{d^4k}{(2\pi)^4}\ 
	\delta \left( p^{*+}x + k^{*+} \right) 
\nonumber \\ && \mbox{}
\hspace{-5ex} 
	\times \mathrm{Tr} \left[ \gamma_5\lambda_\alpha \,S^{*}(k^*)\,\gamma^+
	\hat{P}_q\,S^{*}(k^*)\,\gamma_5\lambda_\alpha^\dagger\,S^{*}(k^*+p^*) \right].
\end{eqnarray}   
It should be noted that the Bjorken-$x$ variable appearing in the above equations is defined in nuclear medium.

To evaluate these quantities we first take the moments defined by
\begin{align}
\label{eq:valence2}
\mathcal{A}_n &= \int_0^1 dx\, x^{n-1}\, q(x),
\end{align}
where $n = 1,\,2,\dots$ is an integer. 
Using the Ward-like identity, $S(k) \gamma^+ S(k) = -\partial S(k)/\partial k_+$, and the Feynman parametrization, 
the quark and antiquark distributions for the $K^+$-meson ($u\bar s$) are obtained as
\begin{eqnarray}
\label{eq:valence51}
u_{K^+}^{} (x)  &=& \frac{3\,g_{K q q}^{*2}}{4\pi^2}  \int d\tau\
e^{-\tau \left[ x(x - 1)\,m_K^{* 2} + x\,M_s^{* 2} + (1-x)\,M_l^{* 2} \right]} 
\nonumber \\ && \mbox{}
\times \left[\frac{1}{\tau} + x(1 - x)\left[m_K^{* 2} - (M_l^{*} 
	- M_s^{*})^2\right] \right], \\ 
\label{eq:valence52}
\bar{s}_{K^+}^{}(x)  &=& \frac{3\,g_{K q q}^{*2}}{4\pi^2}  \int d\tau\
e^{-\tau \left[ x(x - 1)\,m_K^{* 2} + x\,M_l^{* 2} +  (1-x)\,M_s^{* 2} \right]} 
\nonumber \\ && \mbox{}
\times \left[\frac{1}{\tau} + x(1 - x)
\left[m_K^{* 2} - (M_l^{*} - M_s^{*})^2\right]\right].
\end{eqnarray}
Valence-quark distributions of the $\pi^+$ can be obtained by replacing $M_s^{*} \to M_l^{*}$ and $g_{K q q}^{*} \to g_{\pi q q}^{*}$, 
which leads to $u_{\pi^+}(x) = \bar{d}_{\pi^+}(x)$. 
The in-medium valence-quark distributions of other pseudoscalar mesons can be related by flavor symmetry.
The expressions given by Eqs.~(\ref{eq:valence51}) and (\ref{eq:valence52}) are consistent with those given in 
Ref.~\cite{HCT16} for zero baryon density.

The valence-quark distributions in medium and in vacuum, with the corresponding Bjorken variables 
$\tilde{x}_{a}$ and $x_a$, respectively, are related by~\cite{MBITY03}
\begin{align}
  \label{eq:valence6a}
  q_{K^{+}}^{} (x_a) &= \frac{\epsilon_F^{}}{E_F} q_{K^{+}}^{*} ( \tilde{x}_a ), 
\end{align}
with
\begin{equation}
\tilde{x}_a = \frac{\epsilon_F^{}}{E_F} x_a - \frac{V^0}{E_F}
\end{equation}
where $\epsilon_F^{} = \sqrt{(k_F^{q})^{2} + (M_q^{*})^{2}} + V^0 \equiv E_F + V^0$, with $V^0$ being the vector potential, 
is the in-medium quark energy ($\epsilon_F^{} = E_F - V^0$ for an antiquark) and $k_F^q$ is the quark Fermi momentum 
which is related to the nuclear matter density as $\rho_B^{} = 2(k_F^{q})^{3}/3\pi^2$. 
The above formulas are valid only for light ($u$, $d$) quarks in the present approach.
The values obtained by the QMC model for these quantities for the light quark are given in Table~\ref{tab:model3}.

\begin{table}[t]
\caption{Fermi momentum $k_F^q$ and the time component of the vector potential ($V^0$) for the light quark obtained 
in the QMC model for various values of ($\rho_{B}^{}/\rho_0$) with the quantities given in Table~\ref{tab:model1} for 
$m_l=16.4$~MeV.
$k_F$ and $V^0$ are given in units of MeV.}
\label{tab:model3}
\addtolength{\tabcolsep}{20.4pt}
\begin{tabular}{ccc} 
\hline \hline
$\rho_B^{} / \rho_0^{}$ & $k_F^q$ & $V^0$ \\[0.2em] 
\hline
$0.00$   & $0$ & $0$ \\ 
$0.25$   & $162.18$ & $11$ \\
$0.50$   & $204.33$ & $21$ \\
$0.75$   & $233.90$ & $32$ \\
$1.00$   & $257.44$ & $43$ \\
$1.25$   & $277.32$ & $53$ 
\\ \hline \hline
\end{tabular}
\end{table}

The in-medium valence-quark distributions satisfy the baryon number and momentum sum rules as
\begin{align}
\label{eq:valence6}
& \int_0^1\!\! dx \left[ u_{K^{+}}^{} (x) - \bar{u}_{K^+}^{} (x) \right] = 1 , \nonumber \\
& \int_0^1\!\! dx \left[ \bar{s}_{K^{+}}^{} (x) - s_{K^+}^{} (x) \right] = 1, \nonumber \\
& \int_0^1 dx\ x \left[ u_{K^+}^{}(x) + \bar{u}_{K^+}^{}(x) 
	+ s_{K^+}^{} (x) + \bar{s}_{K^+}^{} (x) \right] = 1. 
\end{align}
for the $K^+$. 
Analogous relations hold for the $\pi^+$ as well.

\begin{figure*}[t]
\centering\includegraphics[width=\textwidth]{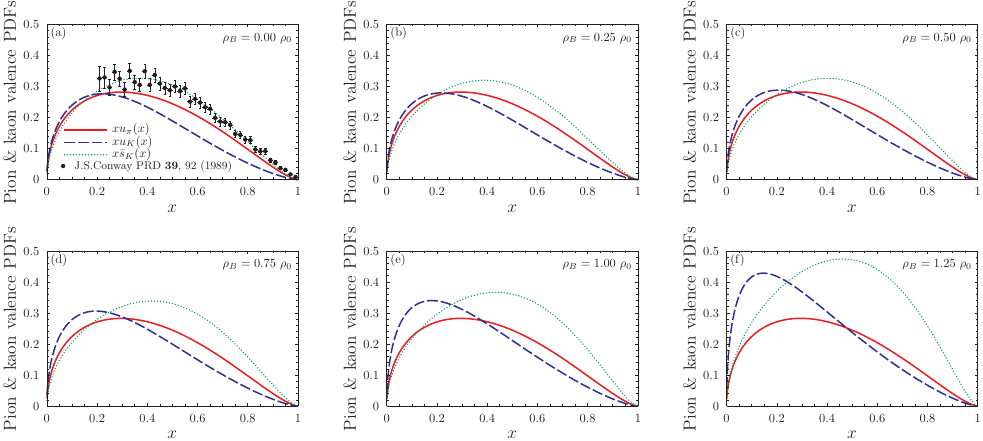}
\caption{ \label{fig5} 
Valence-parton distribution functions of $\pi^+$ and $K^+$ mesons in vacuum and in symmetric nuclear matter, 
evolved from the model scale of $Q_{0}^{2}=0.16~\textrm{GeV}^{2}$ to $Q^2 = 16~\mbox{GeV}^2$
using the NLO DGLAP evolution equations.
(a) $\rho_B^{}/\rho_0^{} = 0.0$, (b) $0.25$, (c) $0.5$, (d) $0.75$, 
(e) $1.0$, and (f) $1.25$. 
The solid lines represent the light ($u$ or $\bar{d}$) quark distributions of the $\pi^+$ and the dotted and dashed lines are those 
for the $\bar{s}$ and $u$ quarks of the $K^{+}$ meson. 
The experimental data for the $u$-quark distribution of the $\pi^+$ are from Ref.~\cite{CAAA89}.}
\end{figure*}

\begin{figure*}[t]
\centering\includegraphics[width=\textwidth]{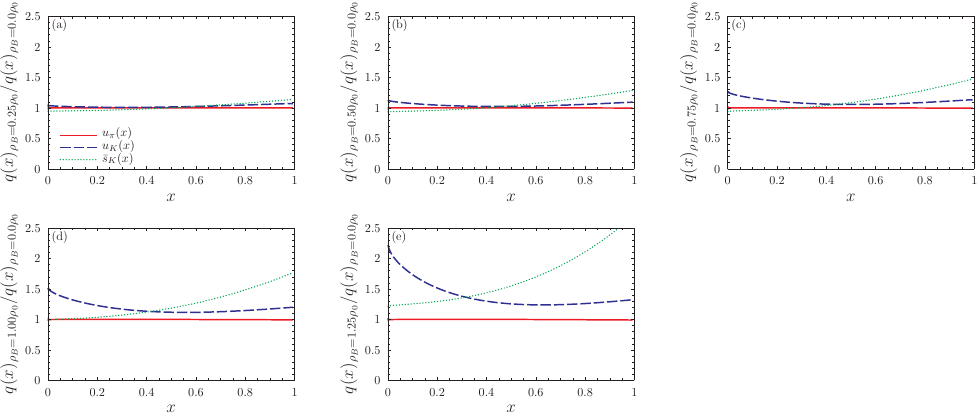}
\caption{ \label{fig6} 
Same as in Fig.~\ref{fig5} but for the ratios of the valence quark distributions of the $\pi^+$ and $K^+$ to the vacuum valence quark 
distributions.
The solid, dashed, and dotted lines are for $u^*_\pi(x) / u_\pi(x)$, $u_K^*(x)/u_K(x)$, and $\bar{s}^*_K(x)/\bar{s}_K(x)$,
respectively.
(a) $\rho_B^{}/\rho_0^{} = 0.25$, (b) $0.50$, (c) $0.75$, (d) $1.00$, and (e) $1.25$. 
In all cases the results have been evolved 
}
\end{figure*}

\begin{figure*}[t]
\centering\includegraphics[width=\textwidth]{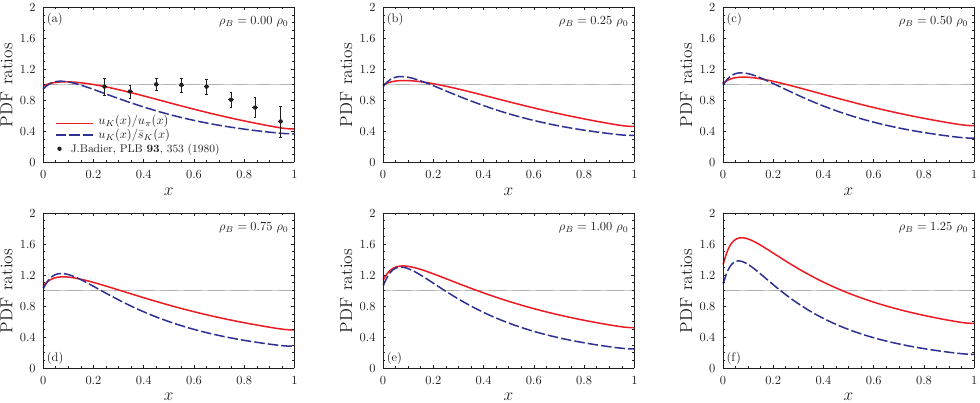}
\caption{ \label{fig7} Ratios of the quark distributions for several densities. 
The solid lines are ratios of the $u$-quark distribution of the $K^+$ to that of the $u$-quark distribution of the $\pi^+$, after the NLO evolution to $Q^2
=16~\mbox{GeV}^2$, while those of the $u$ to $s$ quark distributions of the $K^+$ are represented by the dashed lines. 
The experimental data are from Ref.~\cite{Saclay-80}.}
\end{figure*}

The results for the valence PDFs of the $\pi^+$ and $K^+$ mesons in symmetric nuclear matter are presented in Figs.~\ref{fig5}--\ref{fig7} 
together with those in vacuum. 
The valence-quark distributions have been evolved using the next-to-leading order (NLO) Dokshitzer-Gribov-Lipatov-Altarelli-Parisi (DGLAP) 
evolution equations~\cite{MK95,GL72,AP77,Dokshitzer77} from the model scale of $Q_0^2 = 0.16~\mbox{GeV}^2$, 
which was determined in Ref.~\cite{CBT05} as a typical of valence-dominated models for studying nucleon valence-quark distributions, 
to $Q^2 = 16~\mbox{GeV}^2$. 
The $Q^2$ evolution is carried out in order to compare 
with the experimental data available at $Q^2 = 16$~GeV$^{2}$.

Shown in Fig.~\ref{fig5} are the results for the valence $u$-quark distributions of the $\pi^+$ and $K^+$ and the valence $s$-quark distribution of 
the $K^+$ at $Q^2 = 16~\mbox{GeV}^2$.
The results are shown in vacuum and for symmetric nuclear matter at $\rho_B^{} / \rho_0^{} = 0.25$, $0.5$, $0.75$, $1.0$, and $1.25$. 
In Fig.~\ref{fig5}(a) the obtained valence $u$-quark distributions of the $\pi^+$ are compared with the empirical data of Ref.~\cite{CAAA89}.
Although our results underestimate the available experimental data by up to 20\%, the general behavior of the valence PDF is in reasonable 
agreement with the data.
Comparison of the valence PDFs for several nuclear matter densities shows that the density dependence of the light-quark PDFs of the $\pi^+$ is
rather mild, which is consistent with the conclusion of Ref.~\cite{Suzuki95}.
This can also be verified in Fig.~\ref{fig6} which shows the in-medium to in-vacuum ratios of the valence PDFs.

However, the valence PDFs of the $K^+$ show a different density dependence as shown by the dashed and dotted lines in Fig.~\ref{fig5}.
Not only the magnitude but the shape of the valence PDFs of the $K^+$ change with densities.
In particular, the peak position of $x \bar{s}(x)$ in Fig.~\ref{fig5} changes 
from $x \approx 0.37$ in vacuum to $x \approx 0.45$ at $\rho_B^{} = \rho_0^{}$.  
The change can easily be verified in Fig.~\ref{fig6}, which shows that the valence $u$-quark distribution of the $K^+$ changes noticeably, 
in particular, in the small-$x$ region. 
This feature becomes remarkable at higher densities.
The enhancement is almost 50\% in the small-$x$ region at normal nuclear density.
In contrast to the enhancement of the valence $u$-quark distribution of the $K^+$ in the small-$x$ region, the valence $\bar{s}$-quark distribution 
of the $K^+$ is mostly enhanced in the large-$x$ region as density increases.
This enhancement is even larger than that of the valence-$u$-quark distribution 
in the small-$x$ region, and experimental measurements are highly required to 
verify this prediction.

Finally, in Fig.~\ref{fig7} we present relative strength of the $u$-quark distribution of the $K^+$ with respect to the other quark distributions,
i.e., $u_K(x)/u_\pi(x)$ by the solid lines and $u_K(x)/\bar{s}_K(x)$ by the dashed lines.
Note that all the distributions in Fig.~\ref{fig7} are those of quarks (or partons), but not 
the valence ones. 
Our results for  $u_K^{}(x)/u_\pi^{}(x)$ in vacuum is compared with the available 
experimental data of Ref.~\cite{Saclay-80} in Fig.~\ref{fig7}(a).
One can find that this ratio is enhanced at higher densities and, in particular, in the small-$x$ region. 
In the large-$x$ region, the ratio $u_K^{}(x) / \bar{s}_K^{} (x)$ is suppressed when nuclear density increases. 
This behavior is opposite to the case of $u_K^{}(x)/u_\pi^{}(x)$. 
The deviations of these ratios from unity show the pattern of the flavor symmetry breaking. 
Our results demonstrate that the flavor symmetry breaking effects become larger as density increases.

\section{Summary} \label{summary}

To summarize, we have studied the current-quark properties in symmetric nuclear matter in the QMC model.
Then, using the in-medium quark properties obtained in the QMC model as inputs, we have studied the in-medium properties of dynamical quarks, 
pions, and kaons in symmetric nuclear matter in the NJL model.
In particular, the valence-parton (valence-quark) distribution functions of $\pi^+$ and $K^+$ mesons in symmetric nuclear matter 
in the NJL model were investigated with the proper-time regularization scheme.

In the present study, we estimated the quark condensates, dynamical quark masses, meson decay constants, 
and meson-quark coupling constants for pions and kaons in symmetric nuclear matter in the NJL model.
The valence parton distribution functions of $\pi^+$ and $K^+$ mesons in vacuum as well as in symmetric nuclear matter  
at $Q^2=16~\mbox{GeV}^2$ were calculated and compared with the available data.
We found that the effects of nuclear medium on the valence $u$-quark distribution of the $\pi^{+}$ is rather weak, 
which supports the observation of Ref.~\cite{Suzuki95}.
However, the valence quark distributions of the $K^+$ show appreciable medium effects, namely, the valence $u$-quark distribution 
of the $K^+$ shows enhancement in the small-$x$ region, while that of the $\bar{s}$ shows enhancement in the large-$x$ region. 
The ratios, $u_{K} (x) / u_{\pi} (x)$ and $u_{K} (x) / \bar{s}_{K} (x)$, were found to indicate the flavor symmetry breaking pattern.
This implies that the quark distributions would depend on the surrounding quarks and on the hadron species they reside.
The flavor symmetry breaking effects become larger at higher densities, and the two ratios show the opposite density-dependence 
in the large-$x$ region, although they show a similar density-dependence in the small-$x$ region.
Experimental confirmation is, therefore, highly desired.

For future prospects, a few comments are in order.
First, in the present work, we estimated the in-medium dynamical quark mass. 
This feature can be improved by including momentum-dependent dynamical quark masses generated in nuclear medium, e.g., 
based on the Schwinger-Dyson equations.
Second, it would be interesting to extend the present approach to heavy mesons with charm or bottom flavor. 
These studies would give us further hints on the dynamical chiral symmetry breaking and the realization of heavy quark spin symmetry
through the evolution of heavy systems in nuclear matter.
Moreover, for an alternative elaborated study in the future, it would be interesting to apply a bosonized version of the NJL model
or the NJL model proposed in Ref.~\cite{BT01} to describe the nuclear matter properties, in particular, 
the quark distributions of hadrons in a nuclear medium.

\begin{acknowledgments}
K.T. and J.J.C.M thank the Asia Pacific Center for Theoretical Physics (APCTP) and Kyungpook National University for 
warm hospitality and support during their visits. 
This work was supported by Kyungpook National University Bokhyeon Research Fund, 2017.
P.T.P.H. was supported by the Ministry of Science, ICT and Future Planning, Gyeongsangbuk-do, and Pohang City. 
The work of K.T. was supported by the Conselho Nacional de Desenvolvimento Cient\'{i}fico e Tecnol\'{o}gico - CNPq  
Process, No.~313063/2018-4 and No.~426150/2018-0, and was also part of the projects, Instituto Nacional de Ci\^{e}ncia 
e Tecnologia - Nuclear Physics and Applications (INCT-FNA), Brazil, Process, No. 464898/2014-5, and FAPESP Tem\'{a}tico, Brazil, 
Process. No. 2017/05660-0.
\end{acknowledgments}

\end{document}